\documentclass[conf]{new-aiaa}
\usepackage[utf8]{inputenc}
\usepackage{graphicx}
\usepackage{amsmath}
\usepackage[version=4]{mhchem}
\usepackage{siunitx}
\usepackage{longtable,tabularx}
\usepackage{float}
\usepackage{graphicx}
\usepackage{caption}
\DeclareCaptionLabelFormat{boldlabel}{\textbf{#1~#2}}
\DeclareCaptionFont{myitalic}{\normalfont\itshape}
\title{\textbf{Low-Power Solar Sail Control using In-Plane Forces \\ from Tunable Buckling of Kirigami Films} }

\author{Gulzhan Aldan\footnote{Graduate Student, Department of Mechanical Engineering and Applied Mechanics, AIAA Student Member} and Igor Bargatin\footnote{Associate Professor, Department of Mechanical Engineering and Applied Mechanics, AIAA Member}\footnote{Corresponding Author. Email: bargatin@seas.upenn.edu}}
\affil{University of Pennsylvania, Philadelphia, Pennsylvania, 19104, United States}

\begin{document}

\maketitle

\begin{abstract}

We present a proof-of-concept study showing that buckled aluminized polyimide films perforated with millimeter-scale cuts can redirect normally incident light obliquely and generate net in-plane force components parallel to the global solar sail surface. We use finite element simulations to obtain the buckled shapes of different periodic unit cell geometries and apply ray optics modeling to compute the resulting light-pressure forces. The simulations show that the buckled kirigami surfaces reflect light into different directions producing a net in-plane force parallel to the direction of stretching. We verify these trends experimentally by illuminating a tensioned kirigami sample with a laser and observing reflected beam patterns consistent with the ray optics simulations. These results suggest that kirigami films may offer a scalable, low-power, and lightweight way to achieve controllable in-plane forces for solar sail steering.

\end{abstract}

\section{Introduction}

Solar sailing is a promising propulsion technology for long-distance space exploration missions and interstellar travel \cite{vulpetti2008novel}. While the velocity budget of traditional fuel-based spacecrafts is fundamentally limited by the "tyranny of the rocket equation", solar sails can reach high sustained speeds using nothing but the momentum of light. Since the momentum transfer from a single photon is small, the sail must be lightweight and have a large reflective area to achieve the desired acceleration \cite{montgomery2017}. However, large-area membranes have high moments of inertia and are prone to structural vibrations, which create challenges in realizing attitude control \cite{tsuda2013challenges}.

A number of attitude control methods for solar sails have been proposed in the past. Examples include the use of small reaction wheels and magnetic torquers \cite{polites2008solar}; the tip vane control, where torque is generated through the rotation of vanes \cite{mangus2004solar}; and the sliding mass method, which enables attitude control by adjusting the distance between the sail's center of mass and center of pressure \cite{scholz2011performance}. However, the practical implementation of these methods is limited due to deployment-related issues (tip vanes), limited applicability for interplanetary and interstellar missions (magnetic torquers), or general hardware constraints \cite{wie2007solar}.

In IKAROS, the first and only successful interplanetary solar sail mission to date, attitude control was achieved using Reflectivity Control Devices (RCDs), which were mounted around the perimeter of the sail \cite{tsuda2013achievement, funase2011orbit}. RCDs are devices made of two thin polyimide films, where the bottom film is coated with aluminum and separated from the top film by a layer of the polymer-dispersed liquid crystals. When voltage is applied, the liquid crystals align with the electric field and allow light to pass through and reflect off the planar aluminum coating following the law of reflection. When voltage is not applied, the randomly oriented crystals reflect the incident light diffusely leading to a lower momentum transfer. Selectively activating some RCDs and deactivating others allows adjusting the distribution of the solar radiation pressure across the sail surface and, therefore, its orientation. 

At the same time, resultant non-uniform force components acting normal to the sail surface can also lead to undesirable deformations of the sail membrane \cite{ishida2017optimal}. To address this problem, JAXA developed the Advanced RCD design (A-RCD) \cite{ishida2017optimal}, where the aluminum layer is deposited as a periodic sawtooth structure. Similarly to RCDs, to keep the A-RCDs on, the voltage has to be supplied to them. However, when activated, the sawtooth structure reflects light obliquely generating in-plane force components parallel to the sail surface and reducing the so-called windmill effect. Strategically placed and oriented A-RCDs can produce in-plane forces to achieve the needed torque about the roll axis and enable other sophisticated sail maneuvers.

The idea of using tailored in-plane force components for attitude control has also been recently explored within the metamaterial research community. One of the ways of generating tangential to the surface forces is to engineer optical metamaterials with subwavelength features or gratings that could, for example, scatter incident light into multiple diffraction peaks \cite{swartzlander2017radiation,swartzlander2022theory, dubill2021circumnavigating} or reflect light anomalously \cite{sun2012high, ullery2018strong, joly2023anomalous}. Here we consider the alternative approach of using mechanical metamaterials, in particular, perforated kirigami sheets \cite{rafsanjani2017buckling, bertoldi2017flexible, An2020, Jin2024}. Thin perforated structures buckle under critical tensile strain forming non-planar configurations that change with applied strain until reaching a deformation limit. By tailoring the shape, size, and orientation of the cuts, these in-plane and out-of-plane deformations can be programmed for beam steering and redirection of light across different wavelength for a range of applications \cite{He2024, Liu2023, lamoureux2015dynamic, li2024kirigami, wang2017kirigami} including the potential for solar sail steering, as previously mentioned in the conclusions of Ref. \cite{alderete2021programmable}. 

In this work, we study one possible geometric perforation pattern that can produce in-plane light-pressure forces generated due to the buckling under tension. The presented pattern is designed such that under stretching it forms a globally planar but locally out-of-plane configuration with periodically tilted segments that act like small individual mirrors. The use of oblique reflection to produce in-plane force components is similar to that of A-RCDs, but with the added advantage that the angle of reflection and the resulting in-plane forces can be tuned by controlling the tensile strain applied to the perforated film. Such tuning can be accomplished using, e.g., a stepper electric motor that consumes little electric power when the angle of reflection is changed and zero power when the angle is held constant at a desired value.

The paper is organized as follows. Section~\ref{sec:geometry} introduces the proposed perforation geometry. Section~\ref{sec:simulations} presents the results of a series of tensile and ray optics finite element simulations showing how the reflected ray power and trajectories, generated net in-plane force component, and required actuation force change as a function of strain for tested unit cell designs. Section~\ref{sec:experiments} reports the results of a proof-of-concept experiment that agrees well with simulated predictions and demonstrates the feasibility of the proposed pattern and solar sail control. Finally, Section~\ref{sec:conclusion} concludes the paper and outlines directions for future work.

\section{Geometry}
\label{sec:geometry}

\begin{figure}[H]
    \centering
    \includegraphics[width=0.9\linewidth]{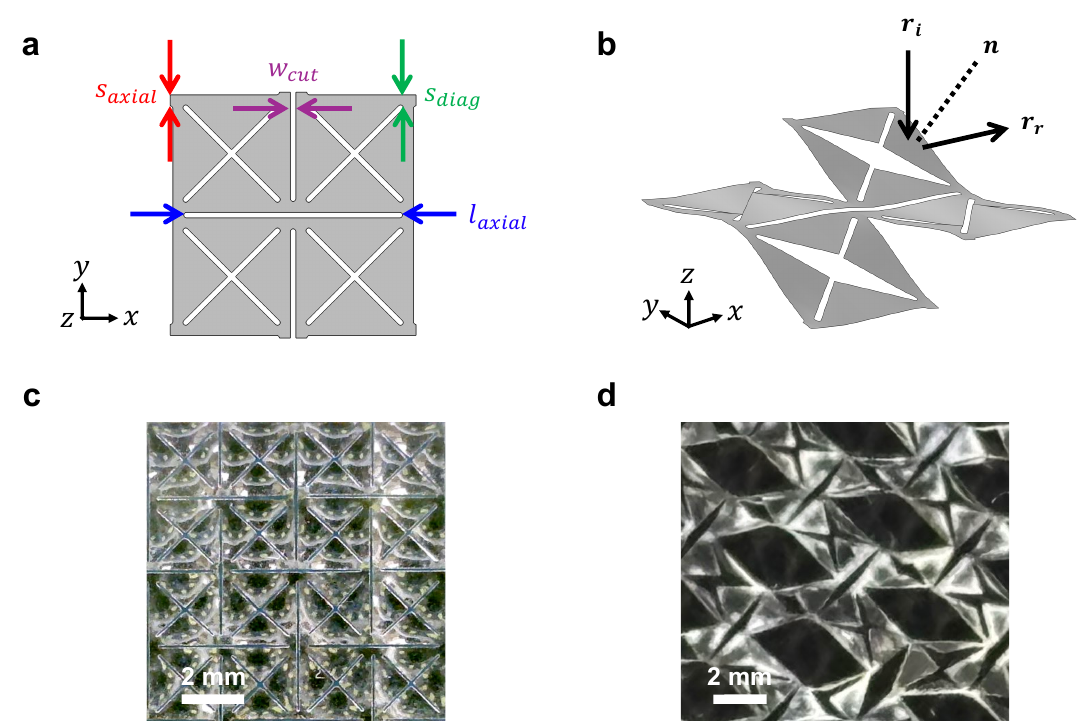}
    \caption {(a) A schematic diagram of the unit cell geometry \cite{aldan2025kirigamifilmreflectordeployable}; (b) an example of the simulated shape of the buckled unit cell under tensile strain applied in the \textit{x}-direction with the dotted line showing the normal to the surface, \(\boldsymbol{n}\), at an arbitrary point on the unit cell, and arrows showing the directions of the incident and reflected rays, $\boldsymbol{r}_i$ and $\boldsymbol{r}_r$, respectively; (c)--(d) the laser-cut 7.8~$\mu\text{m}$ thick aluminized polyimide film in the (c) undeformed and (d) buckled states.}
    \label{fig:1}
\end{figure}

The proposed kirigami sail is a periodic structure consisting of a two-dimensional array of unit cells perforated with axial and diagonal cuts (Fig.~\ref{fig:1}). We recently studied how this pattern affects the mechanical properties of thin aluminized polyimide films in our earlier work~\cite{aldan2025kirigamifilmreflectordeployable}. A key feature of this design is that it is highly flexible and stretchable requiring relatively small forces to reach large strains. In the following sections, we conduct a proof-of-concept study to show that buckled films perforated with this pattern, using millimeter-scale cuts, can redirect normally incident light obliquely and, therefore, generate light-pressure-induced in-plane forces.

\section{Mechanical and ray optics simulations}
\label{sec:simulations}

\begin{figure}[H]
    \centering
    \includegraphics[width=1\linewidth]{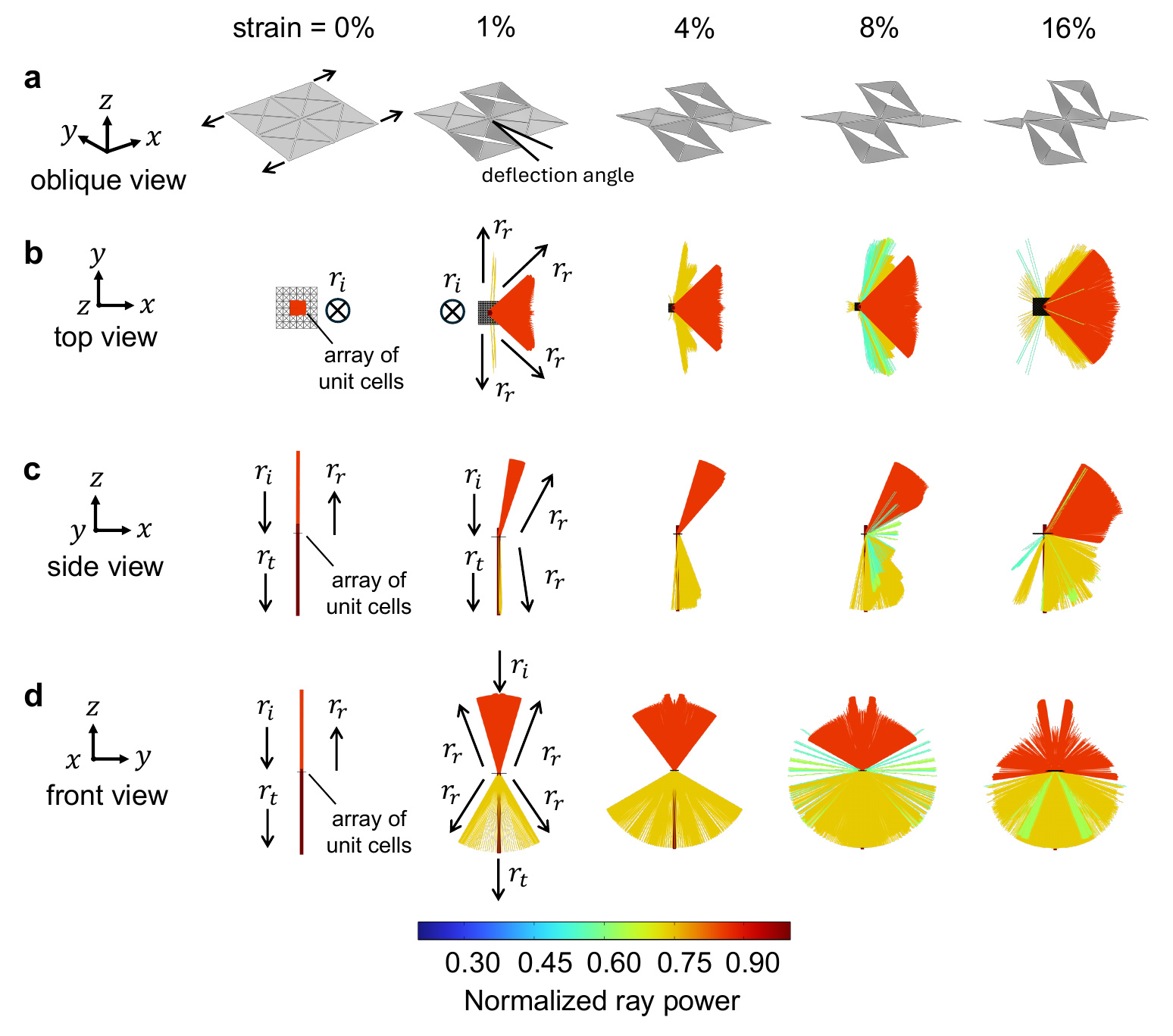}
    \caption {(a) An oblique view on the simulated deformation of the periodic unit cell with $s_{\text{axial}}=s_{\text{diag}}=s=0.3~mm$ and $l_{\text{axial}}=9.3~mm$ under varying strains; (b)--(d) for each corresponding strain, the central unit cell in the periodic array consisting from 3$\times$3 to 7$\times$7 unit cells is illuminated by a light source having the same size as the unit cell. The diagrams show the simulated incident, reflected, and transmitted rays in the $\boldsymbol{r}_i$, $\boldsymbol{r}_r$, and $\boldsymbol{r}_t$ directions, respectively, with color indicating the normalized ray power which is defined as a ratio of the ray power after and before interaction with the unit cell. In all shown views, the light is incident in the negative z-direction and stretching is applied in the x-direction.}
    \label{fig:2}
\end{figure}

To obtain the shape of the perforated film under different tensile strains, we conducted linear buckling and post-buckling simulations for periodic unit cells. Geometric nonlinearity was included in all steps but not material nonlinearity. The perforated polyimide (PI) film aluminized (Al) on two sides was modeled in the composite materials module in COMSOL as a layered shell. Each layer was defined as an isotropic elastic material with a specified thickness ($t_{\text{PI}} = 7.8~\mu\text{m}$, $t_{\text{Al}} = 100~\text{nm}$), Young's modulus ($E_{\text{PI}} = 2.76~\text{GPa}$, $E_{\text{Al}} = 70~\text{GPa}$), and Poisson's ratio ($\nu_{\text{PI}} = 0.34$, $\nu_{\text{Al}} = 0.33$) ~\cite{DuPont2022}. A uniaxial tension was applied to the unit cell in the longitudinal direction along its axial cuts. To enforce periodicity, symmetry boundary conditions were imposed on the lateral edges, and periodic boundary conditions were applied on the edges where tension was prescribed. After the linear buckling analysis, the eigenmode that most closely resembled the experimentally observed deformation was introduced as the initial geometric imperfection for the post-buckling study. 

Fig.~\ref{fig:2}a shows how the deformation of the buckled periodic unit cell changes as the tensile strain increases. The deformation is symmetric about the axial cuts aligned with the stretching direction and anti-symmetric about the cuts oriented perpendicular to it. As a result, the square segments located diagonally opposite each other have similar surface normal vectors. This distribution of surface normals determines how incident light is redirected by the unit cell through the law of reflection. Because of the combined symmetry and anti-symmetry of the deformation, a portion of the normally incident light is expected to be reflected at nonzero angles in at least two distinct directions. The important result to demonstrate next is that the in-plane forces these reflected rays generate do not cancel each other.

The deformed meshes of the buckled unit cells at selected strains were exported from the mechanical simulations and imported into a ray optics model in COMSOL. In the ray optics simulations, the buckled unit cell surfaces were defined as specularly reflective with a solar absorptance of $\alpha = 0.14$ (a maximum reported solar absorptance for the experimentally used films~\cite{Sheldahl2020}) using the wall boundary condition. Although aluminized polyimide reflects light both specularly and diffusely, it was treated as purely specular here for simplicity. This assumption is supported by reported specularity values ranging from 94\% to 98\% for smooth unwrinkled aluminized polyimide films~\cite{rowe1978sail, fieseler2015critical, ralph2000g}. The unit cell was illuminated normally with rays released from a rectangular light source of the same size as the unit cell. The total incident power was set to 1~W, with each ray assigned equal initial power. In the simulations, when a ray hit the unit cell surface, the power of the reflected beam was automatically calculated using the Fresnel equations \cite{comsol}.

Fig.~\ref{fig:2}b--d show how the simulated ray power and trajectories change with increasing tensile strain when viewed from different perspectives. As the film is stretched in the $x$-direction, the reflected rays become more dispersed in all three projection planes, but the diagrams provide qualitative information about the force components that can be generated. First, the distribution of reflected rays is symmetric about the central axial cut aligned with the stretching direction. This symmetry agrees with the symmetry of the unit cell deformation and implies that the net force in the $y$-direction should be zero. Second, the distribution is asymmetric with respect to the direction perpendicular to stretching, which indicates the presence of a net force component in the $x$-direction. Lastly, majority of the rays in the upper half-space result from a single reflection and therefore retain 86\% of their initial power, whereas rays in the lower half-space lose more power due to multiple reflections.

To validate these qualitative observations, we calculated the forces generated by the light pressure. For this purpose, the unit cell was assigned three accumulators corresponding to each force component. Following the momentum conservation and the law of reflection, the force vector associated with each ray was calculated using the accumulator as 
\(\boldsymbol{F} = \frac{P}{c}(\boldsymbol{r}_i - \boldsymbol{r}_r)
= \frac{P}{c}[(1 - R)\boldsymbol{r}_i + 2R(\boldsymbol{r}_i \cdot \boldsymbol{n})\boldsymbol{n}]\),
where \(\boldsymbol{n}\) is the surface normal defined by the buckled unit cell geometry, $\boldsymbol{r}_i$ is the direction of the incident ray, $\boldsymbol{r}_r$ is the direction of the reflected ray, $R$ is the reflectance of the material defined as $1 - \alpha$, $P$ is the ray power, and $c$ is the speed of light in vacuum. The net force components generated for the entire unit cell were then obtained by summing the contributions of all simulated rays.

To ensure the accuracy of the numerically simulated forces, three types of convergence studies were performed. First, to capture the effect of the complex unit cell geometry, the number of rays released from the light source was chosen based on how the calculated forces converged with increasing ray count. Between 10000 and 30000 rays were sufficient for the geometries tested in this work. Second, due to the complex shape of the buckled unit cells, multiple reflections can occur, which may reduce the net in-plane force compared to an isolated illuminated unit cell with no neighboring cells. To capture this effect, the perforated film was modeled as a periodic array of unit cells with increasing size from $1\times1$ to $7\times7$ until the calculated forces converged. Typically, arrays of $3\times3$ or $5\times5$ unit cells were sufficient to capture the effect of multiple reflections. Lastly, a mesh-refinement convergence study was also performed.

\begin{figure}[H]
    \centering
    \includegraphics[width=1\linewidth]{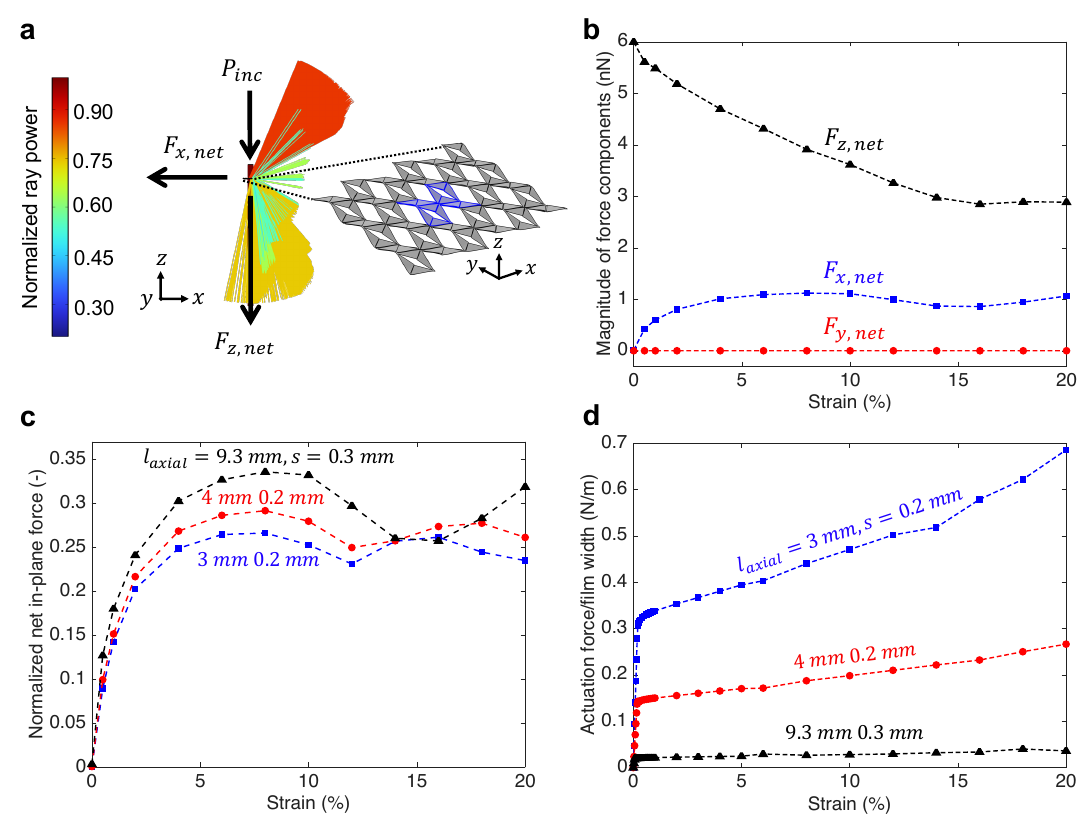}
    \caption {(a) A side view on the simulated ray trajectories with arrows showing the directions of the generated net in-plane component of the force, $F_{\text{x, net}}$, and net out-of-plane component of the force, $F_{\text{z, net}}$, due to light with the total power of $P_{\text{inc}}$, and the inset enlarging a simulated array of periodic unit cells located in the xy-plane with $l_{\text{axial}}=9.3~mm$ and $s=0.3~mm$ under 8\% tensile strain applied in the \textit{x}-direction; (b) the magnitude of the simulated generated net force components as a function of strain for the unit cell in (a) when $P_{\text{inc}}=1~W$; (c) simulated normalized net in-plane force defined as $F_{\textit{in-plane, norm}}=cF_{\textit{x, net}}/P_{\textit{inc}}$ for stretched unit cells with varying geometric parameters $l_{\text{axial}}=9.3~mm$ and $s=0.3~mm$ (black), $l_{\text{axial}}=4~mm$ and $s=0.2~mm$ (red), and $l_{\text{axial}}=3~mm$ and $s=0.3~mm$ (blue); (d) corresponding simulated actuation force divided by the film width as a function of strain for periodic unit cells in (c).}
    \label{fig:3}
\end{figure}

The force simulations confirm that a net in-plane force exists and it acts in the direction of the applied stretching and is on the order of a nanoNewton for a unit cell illuminated with 1~W of power (Fig.~\ref{fig:3}a--b). At the same time, the net $y$-component of the force remains zero for all tensile strains. There is also a net out-of-plane force acting in the direction of the incident light, which is on the order of several nanoNewtons per 1~W of illumination. The tunability of the out-of-plane component may also be useful for inducing yaw and pitch rotations of the sail.

For each simulated case, the in-plane normalized force was defined as \(F_{\text{in-plane,norm}} = c\,F_{x,\text{net}} / P_{\text{inc}}\), where \(F_{x,\text{net}}\) is the net in-plane force component in the stretching direction and \(P_{\text{inc}}\) is the total incident power. The results of the buckling and ray optics simulations for unit cells with axial cut lengths of 3, 4, and 9.3~mm stretched up to 20\% strains are summarized in Fig.~\ref{fig:3}c. Regardless of cut length, the net in-plane force gradually increases with strain, reaches a maximum at around 8\% tensile strain, then slightly decreases up to about 12\% strain, and increases again at higher strains. The simulations also show that larger perforation sizes yield higher in-plane forces with maximum normalized forces of around 0.34, 0.29, 0.27 for the 9.3~mm, 4~mm, and 3~mm perforations, respectively. 

These values are not small and comparable to the theoretical maximum allowed for the rotating vanes assuming single reflections only. For comparison, the maximum normalized in-plane force produced by a solid mirror of reflectance $R$ illuminated under constant light intensity is $4R/(3\sqrt{3})\approx 0.77 R\approx 0.66$ at the tilting angle of  $\arcsin(1/\sqrt{3})\approx$35.5\textdegree. Our stretched kirigami film therefore produces about one half of the force of an optimally tilted mirror of the same area but without the need for rotation/tilting. For very large solar sails, the moment of inertia of rotating vanes may become prohibitively large for quick reorientation, potentially creating an advantage for our structure, which produces a twice smaller in-plane force but can be actuated more quickly and without exciting strong vibrations in the main sail. Exploring other geometric parameters or alternative unit cell designs could help determine whether even larger normalized forces are achievable.

The corresponding tensile forces required to reach the simulated strains are also summarized in Fig.~\ref{fig:3}d, and show that the current optimum geometry with 9.3~mm cuts requires forces of around 0.03~N per meter of width of the film, which implies relatively low force and power requirements for actuation. In a solar sail spacecraft, these deformations could be induced by thin, lightweight cables that, when pulled, locally stretch the sail and produce the desired shape change. With multiple cables and pulleys, different regions of the sail could be stretched independently, providing real-time control of the magnitude and spatial distribution of the in-plane forces. This offers a simple and mechanically tunable alternative to conventional roll and attitude control systems.

\section{Experimental observation of oblique reflection}
\label{sec:experiments}

\begin{figure}[H]
    \centering
    \includegraphics[width=1\linewidth]{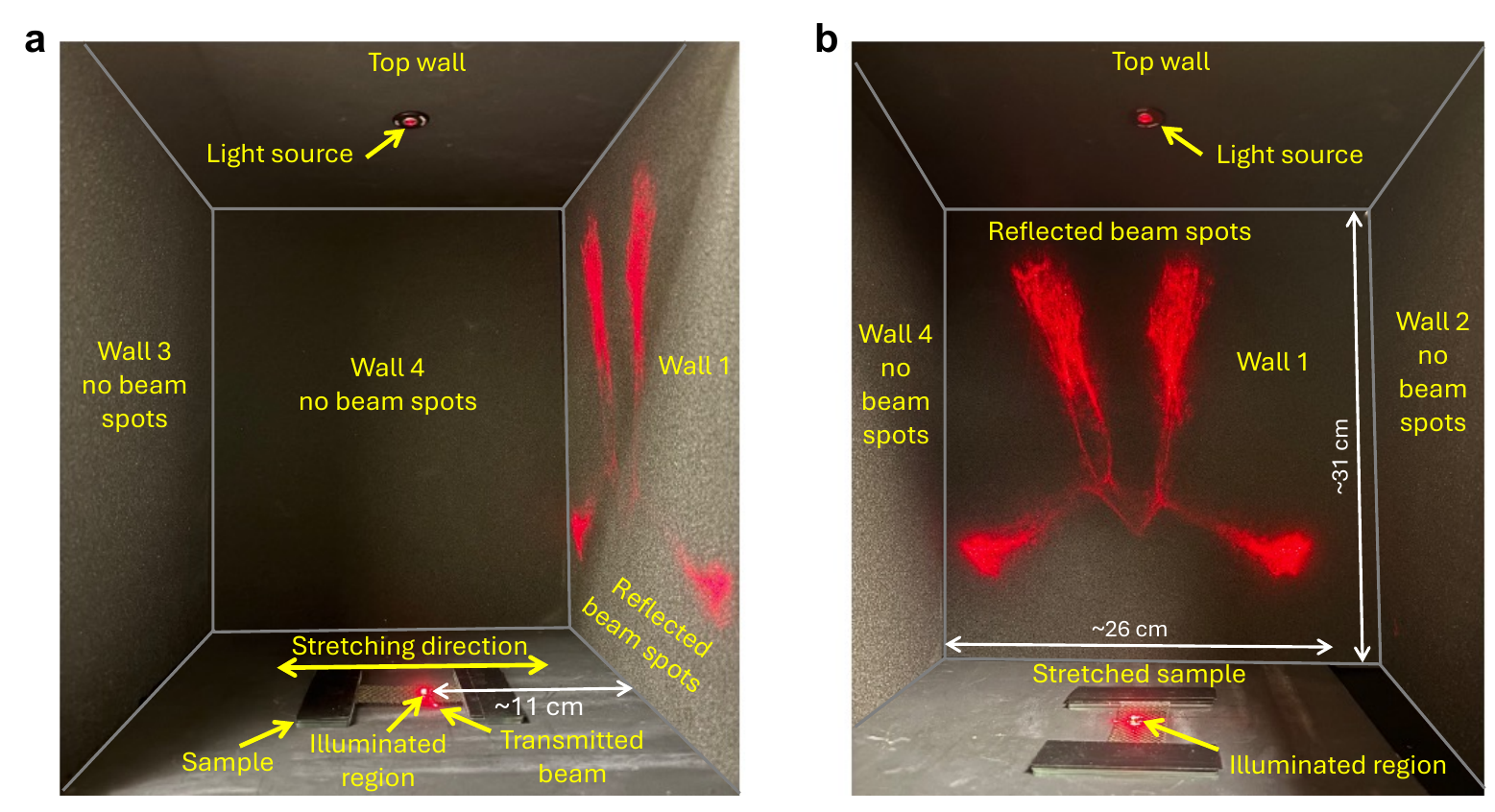}
    \caption {Photographs of the (a) side view on the experimental setup from the wall 2 perspective showing the normally illuminated region on the stretched sample, transmitted beams, and reflected beam spots located only on wall 1; and (b) front view from the wall 3 perspective showing the shape of reflected beam spots on the wall 1 and no beam spots on wall 2. The wall 1 was $\sim$26~cm wide, $\sim$31~cm tall, and $\sim$11~cm away from the illuminated region on the sample. Note: the grey edges at the intersections between adjacent walls were added manually to make the walls easier to distinguish.}
    \label{fig:4}
\end{figure}

To validate the simulated predictions about the direction of the generated forces, we conducted a qualitative proof-of-concept experiment to show that light redirected obliquely from a buckled perforated aluminized film can produce a net nonzero in-plane force. The experimental setup consisted of a light source, a kirigami sample, and a compartment with non-reflective walls surrounding the sample. The compartment was formed by four planar black polyurethane foam vertical walls, which do not re-reflect the incident beam, and black-tinted MDF boards at the top and bottom. For reference, we labeled the vertical walls as wall~1 through wall~4 (Fig.~\ref{fig:4}).

The tested sample was a $7.8~\mu\text{m}$ thick aluminized polyimide film with 3~mm axial cuts and 0.08~mm spacings (bulk uncut film with an areal density of $11~g/m^2$ supplied by Sheldahl Inc.). The cut size was chosen only to demonstrate beam redirection and to ensure that at least one half of a unit cell could be illuminated given the spot size of the used light source. The perforated sample was 50~mm~$\times$~26~mm in size and was attached on two opposite sides to glass slides to ensure uniform stretching in the longitudinal direction. The sample was placed on the bottom board of the compartment and stretched along the direction parallel to walls~2 and~4. A 635~nm 4.5~mW laser mounted on the top board above the sample was used as the light source. The laser beam had a rectangular beam spot and normally illuminated half of a unit cell such that its shorter dimension aligned with the stretching direction. Because the unit cells are diagonally symmetric, illuminating half a unit cell was sufficient to capture the redirection behavior of the entire unit cell.

We began the experiment by illuminating the moderately stretched film with the laser. The tilted segments in the buckled pattern redirected the incident beam into distinct directions. As shown in Fig.~\ref{fig:4}, the reflected beam spots appeared only on wall~1. These reflected directions have non-collinear in-plane components in the stretching direction, so taken together they imply a net nonzero in-plane force. No beam spots were observed on walls~2, 3, or~4, indicating that no forces are generated in directions opposite to the beams hitting wall~1. Some light was also transmitted through the kirigami film. Overall, these observations are in good qualitative agreement with the ray optics simulations.

We also note that buckling of the perforated film can in principle happen in one of two ways due to the symmetry of the deformation. The direction the reflected beams is in general randomly chosen to be either toward wall 1 or wall 3 but only during the first test. After the first buckling, the film becomes slightly plastically deformed, and buckles consistently in the same direction afterwards. 

\begin{figure}[H]
    \centering
    \includegraphics[width=0.97\linewidth]{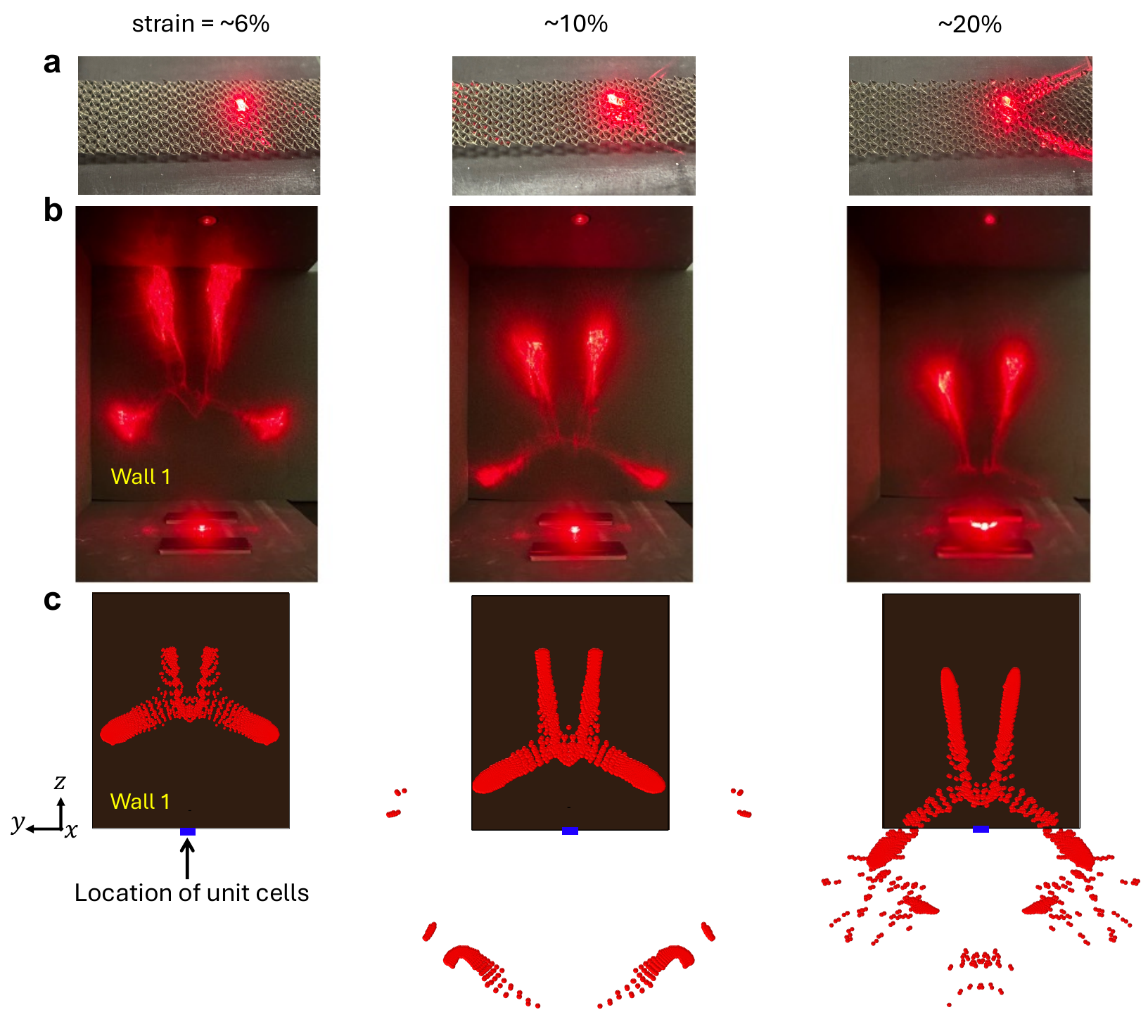}
    \caption {(a) Photographs of the perforated 7.8~$\mu\text{m}$ thick aluminized polyimide film with $l_{\text{axial}}=3~mm$ and $s=0.08~mm$ under varying tensile strains applied in the horizontal direction with normally illuminated half-unit cell; (b) photographs showing the shape and location of the reflected beam spots on the wall 1 under the same strains; (c) corresponding simulated beam spots with the blue marks indicating the location of the array of buckled unit cells at the same strains and the black rectangles representing the simulated wall 1 having the same size and located at the same distance from the illuminated spot as in the experiments. Note that in (c) at 10\% and 20\% strain, the simulations indicate additional beam spots located in the same plane as wall 1 but bellow the sample. These spots were not captured in the experimental photographs shown in (b) because the sample was positioned directly on a solid board tinted black.}
    \label{fig:5}
\end{figure}

Next, we gradually stretched the film further to observe how the shape and location of the reflected beam changed on wall~1. The film was stretched uniaxially from 6\% to 20\% strain while the position of the illumination source remained fixed. A close look at film (Fig.~\ref{fig:5}a) confirmed that the deformation was periodic with clearly defined tilting of the film segments. As the strain increased, the tilt angles changed, which shifted both the shape and the position of the reflected beam spots. With increasing strain, the spots moved downward on wall~1 (Fig.~\ref{fig:5}b). We also observed that the reflected beam pattern remained vertically symmetric, which is consistent with the symmetry predicted in the ray trajectory simulations.

For comparison, we simulated the experimental setup in COMSOL by placing the stretched sample, at the same strains as in the experiment, 11~cm away from a plane representing wall~1. In the simulation, the perforated sample was modeled as a $5 \times 5$ array of buckled unit cells with the same geometric parameters as the experimental sample. The diagrams in Fig.~\ref{fig:5}a show the simulated reflected rays interacting with this plane, where black rectangles denote the wall~1 (having the same size as the experimental wall 1), and blue ticks indicate the position of the simulated sample.

At 6\% strain, the simulated beam spot is slightly shorter than the one observed experimentally. However, the lower portion of the spot and its location agree well with the measurements. The small discrepancy near the top may be due to manual stretching and slight differences between the simulated and actual shapes at this strain. At 10\% and 20\% strain, both the overall beam shapes and their positions show good agreement between simulation and experiment. The simulations also reveal additional beam spots below the unit cell plane at these strains likely arising from multiple reflections within the unit cell geometry. These spots were not observed experimentally because of the limited compartment size and because the perforated sample was suspended on a solid board that did not transmit light.

Overall, the ray optics simulations agree well with the experimental observations demonstrating that the mechanically deformed shape observed in the experiment is reasonably close to the simulated one despite imperfections from manufacturing and the stretching method. Moreover, the similarity in the beam spot locations between experiment and simulation indicates that the ray optics model also provides reasonable predictions of the direction of the redirect light.

\section{Conclusion and future work}
\label{sec:conclusion}
In this work, we presented a proof-of-concept simulations and an experiment showing one possible highly flexible and stretchable kirigami perforation pattern that can be used for solar sail control. Future work may explore a broader range of kirigami geometries and scales to understand how the in-plane and out-of-plane components of the light-induced forces can be tuned and optimized.

\section*{Acknowledgment}
We thank Zhipeng Lu, Matthew Campbell, and Luke Stoner-Eby for helpful discussions. This work supported by the School of Engineering and Applied Sciences at the University of Pennsylvania.

\end{document}